# Effect of Temperature and Added Salt on a Model Polyzwitterion Polymer in Dilute Solution

Soumik Ghosh[1, 2], Vivek M. Prabhu[1], and Jack F. Douglas[1, *]

[1]Materials Science and Engineering Division, National Institute of Standards and Technology, Gaithersburg, Maryland 20899, USA

[2]Department of Physics, Wesleyan University, Middletown, Connecticut 06459-0155, USA

ABSTRACT

The competitive interactions arising from the proximity of positive and negatively charged groups in the monomers in polyzwitterions (PZ) makes these polymers have properties similar in some ways to intrinsically disordered proteins, RNA, and nearly charge-neutral polyelectrolyte complexes between oppositely charged polyelectrolytes. To gain qualitative insights into this important class of water-soluble polymers, we performed molecular dynamics (MD) simulations of a model PZ chain with an explicit solvent to understand the conformational state in the absence and presence of added salt. We find that isolated PZ chains under salt-free conditions adopt a self-associated state having a weak dependence on temperature. However, the addition of an appreciable amount of NaCl leads to a weakening of the intrachain associations, resulting in chain expansion from a relatively contracted to a more extended configuration, a transition reminiscent of the denaturation of globular proteins and the disassociation of polyelectrolyte complexes. In particular, the hydrodynamic penetration function, the ratio of hydrodynamic radius to the radius of gyration, of the PZ chains under salt free conditions, is found to be more consistent with a randomly branched polymer or 'single chain nanoparticle' than a random coil polymer where the physical cross-links within the chain arise from the dipoles within the chain. The chain conformation not only tends to become larger with the addition of salt but also acquires a more appreciable temperature dependence of its average size with temperature. This swelling trend of PZ molecules with added salt, sometimes referred to as the 'anti-polyelectrolyte effect', is distinct from the normal trend polyelectrolytes, which typically become more contracted with the addition of salt. We also examined the spatial extent of the dynamic hydration layer the PZ chain, both with and without added salt, and the effect of temperature and salt concentration on the water mobility within this layer, and PZ side-chain mobility, whose dynamics is apparently strongly coupled to the surrounding water molecules.

Corresponding authors: soumik.ghosh@nist.gov, jack.douglas@nist.gov

***** Emeritus

## Introduction

Polyzwitterions (PZs) belong to a special class of polymer material in which each monomer unit carries a positive and a negative charge center. Their solubility and conformation in a solution varies greatly, depending on the physicochemical properties. Due to their overall charge neutrality, PZs are used in numerous applications including drug delivery,[1] solid-state batteries[2] and antifouling surfaces.[3,4] Over the past few decades, different kinds of polyzwitterions have been synthesized by controlling the types of functional groups on the zwitterionic unit, length of the unit, as well as the spacing of the units along the backbone. Betaines are a particular class of PZ polymers that include the negatively charged sulphonate, or carboxylate, or phosphate group. Proteins, rich in oppositely charged amino acids, behave like polyzwitterion chains.[5] Sulfobetaines are extensively used in applications requiring anti-fouling surfaces due to their resistance to protein adsorption.[6] The presence of a strong hydration layer is found to be the characteristic of such zwitterionic polymers.[7,8] Although sulfobetaines are often used to prepare hydrogel[9] and polymer brush materials,[5,10] the chain conformation in aqueous solution and the effects of side-chain dynamics on their hydration layer are not completely understood. There is particular interest in how polybetaines and other polyzwitterion chains respond to external stimuli, such as changes in temperature and addition of salts to solution. This aspect of PZs requires attention in connection with many applications[11,12] of this class of materials.

Our simulations focus on investigating a number of singular experimental and theoretical observations on the solution properties of PZ polymers. Such nearly charge-neutral polymers, bearing both positive and negative charge centers on their monomer units, can naturally be expected to exhibit some properties in common with proteins[13] and polyelectrolyte complexes composed of associated positive and negatively charged polyelectrolytes.[14] We briefly compare the configurational properties of our simulated PZ polymers with measurements on intrinsically disordered proteins, RNA and polyelectrolyte complexes, as these polymers are likewise expected to exhibit appreciable intermolecular association. The size of a single PZ chain, in the absence of any salt, are often observed to exhibit little variation in their size with temperature[10] so we vary temperature to see if this trend is observed in our simulations of PZ solutions, and we also consider how the addition of salt alters the dimensions of these polymers and the corresponding temperature dependence of their size. Notably, the absence of any appreciable temperature of the PZ size implies no conventional $\theta$-temperature exists at which chains undergo a coil-to-globule transition when temperature is varied, as in neutral flexible polymers in organic solvents. This situation raises questions about whether it is sensible to describe PZ polymers in terms of the conventional random coil polymer model of polymers in solution or by another model that emphasizes the tendency of the charged units within these polymers to self-associate to form *randomly branched polymers* with physical cross-links as an alternative to the random coil structural paradigm. We then look for structural indicators, measurable from dynamic and static scattering measurements to test whether this alternative model of PZ solutions is supported by our simulations. Finally, the general tendency of the addition of salts to increase PZ solubility rather than decrease solubility as normally found in polyelectrolyte solutions, the so-called anti-polyelectrolyte effect[5,15–17], also motivated our computational investigation. Theoretically, the often observed increase in PZ solubility with the addition of salt has been suggested to arise from a change in PZ conformation arising from change in orientation of the dipoles in the PZ side-groups induced by ion adsorption.[10] Accordingly, we examine dipolar orientation in the side-groups of our simulated PZ polymers to determine if we can find computational evidence supporting the existence of such a transition having its origin in the dipolar orientation in the side-groups. All our simulations are performed with explicit water, which provides a reasonable model for simulating the properties of water and the manifold hydration effects inherent to aqueous solutions of charged polymers and salts.

Surprisingly, though there are numerous previous experimental studies[10,15,17–19] on PZ polymer solutions that have guided our computational study, simulations studies of PZ solutions, especially those based on atomistic models, are sparse. Recently, simulations[5] based on a coarse-grained bead-spring polymer model, initially developed[20] to model PZ 2-methacryloyloxyethyl phosphorylcholine solutions, were carried out to better understand the effect of salts on the dipolar interactions within PZ side-chains, but these simulations do not address the complex hydration effects characteristic of real PZ solutions and must be considered exploratory in nature. All-atom simulations are limited due to their high computational cost and the scarcity of forcefield parameters that faithfully describe polymer-solvent and polymer-salt interactions. Although we do include explicit water and model the interactions of this solvent with ions and the polymer, the modeling is still in its developmental stage so our contribution must also be viewed as exploratory in nature.

The original impetus for our molecular dynamics simulations was the striking observations by Pickett *et al.*,[21] who performed benchmark small-angle neutron scattering (SANS) measurements characterizing the equilibrium structure of poly[3-(acrylamidopropyl-dimethyl-ammonium)-propyl-1-sulfonate] (pAPAPS) in semi-dilute solutions with a small amount of added salt to enhance PZ solubility. This study revealed a weak temperature dependence of the average chain size over a wide temperature range (10 °C to 50 °C) and a 'theta point' falling inside this temperature range defined experimentally by a vanishing second osmotic virial coefficient. Pickett *et al.*[21] also observed a modest increase in the average chain dimensions with addition of 0.1 mol/L NaCl. Evidently, the absence of a coil-globule transition does not preclude the existence of a theta point, an initially puzzling observation that we discuss in the Conclusion section in relation to the association-complex model of PZ molecules in solution. We have also been influenced by the observation by Angelopoulos et al.[22] that the model PZ polymer, poly(2-vinylpyridine-N-oxide), exhibits a curious appreciable decrease of the chain radius of gyration ($R_g$) in salt solutions by SANS with increasing temperature around room temperature, even though there is no evidence of incipient phase separation in this system. This is a singular observation that deserves further study.

In the present work, we investigate changes in conformation and dynamics of a single pAPAPS chain by molecular dynamics (MD) simulation with explicit water arising from changes in temperature in both pure water and a 0.5 mol/L NaCl solution. Given the suggestion that dipolar orientation of the sidechains is important in these molecules,[23,24] we quantify the temperature and salt dependence of the orientation of the dipoles present in the side-chains, as well as the distribution of the functional groups around the backbone of the polymer. We also quantify changes in mobility within the hydration layer that 'dresses' the PZ chains and changes in the compactness and average size of the PZ polymers under both salt-free and added salt conditions, and over a range of temperature.

## Methodology

We build our all-atom sulfobetaine chain, pAPAPS, based on the optimized potential, for liquid simulations (OPLS-AA) generalized forcefield (FF)[23,24] with an extension of sulfonate parameters; those parameters are obtained from Lopes *et al.*[25] We use Moltemplate package[26] to create a polymer chain with 50 repeating monomer units. The polymer chain is then solvated with 30000 water molecules and 300 Na − Cl salt pairs when needed. This corresponds to a salt concentration of ≈ 0.5 mol/L in aqueous solution. We use the non-polarizable TIP4P/2005[27] model for water and apply salt parameters from Li *et al.*[28] Geometric combing rules, consistent with OPLS FF, for all cross pairs are employed by common convention.[23,24] We perform NPT simulations with LAMMPS[29], using Nose-Hoover thermo- and barostat.[30] Starting from a rod-like configuration, we perform equilibration for at least 40 ns before performing production runs. Production runs up to 200 ns were performed to gather structural data. For long-range correction to Coulombic interaction, we use the particle-particle particle-mesh (PPPM) algorithm and truncate Lennard-Jones (LJ) interactions at 1.2 nm.

The size of the chain is measured by calculating the radius of gyration, $R_g$, given by,[31]

$$R_g^2 = \frac{1}{m_i}\sum m_i(r_i - r_{CM})^2, \quad (1)$$

where $m_i$ and $r_i$ are the mass and position of the $i^{th}$ atom, and $r_{CM}$ is the center of mass location. We consider a standard metric quantifying the overall 'shape' and compactness[32] of polymer chains in solution, the ratio of $R_g$ and $R_h$, where $R_h$ is the hydrodynamic radius of the polymer. In particular, the hydrodynamic penetration function,[33,34] $\psi_h = R_h/R_g$ approaches a well-known limiting value of $\psi_h(\text{sphere}) = (5/3)^{1/2} \approx 1.29$ for a uniform sphere and a value near 0.8 for relatively diffuse random coil[35–37] polymers, while $\psi_h$ approaches 0 for infinitely long rods.[36,38] The term hydrodynamic penetration function arises from the fact that hydrodynamic interactions are large, and the polymer is correspondingly impenetrable to the solvent velocity field, when $\psi_h$ is relatively large, and $\psi_h$ becomes small when the interior of the polymer is significantly exposed to the solvent velocity field, a situation that has historically been termed 'free draining'. For star polymers, with a moderate number of arms and other appreciably branched polymers, such as dendrimers and randomly branched polymers, $\psi_h$ tends to be closer to 1 or takes even somewhat larger values[39,40] that are more reminiscent of compact particles, as discussed in detail below. $\psi_h$ thus provides valuable experimental information about the relative 'compactification' of polymers relative to a linear random coil reference state in systems undergoing a propensity to self-associate or to mutually associate[41] in polymers which have complementary interactions that lead to complexes between polymers[42]. The value of $\psi_h$ notably also depends somewhat on molar mass (degree of polymerization) of the chain[43,44] and, as noted above, tends to decrease to values well below the random coil value when polymers are highly elongated and to somewhat lower values ($\psi_h = 0.7$) in good solvents where they are somewhat swollen relative to ideal 'random coil' polymers modeled approximately as random walk chains.[36,45] The software routine, ZENO[46] [This software was developed by employees of the National Institute of Standards and Technology (NIST), an agency of the Federal Government and is being made available as a public service.], was used to calculate the hydrodynamic radii, $R_h$, based on chain conformations sampled from our MD simulations. This software uses a novel path-integration methodology in which the polymer chains are probed with random walk trajectories to solve the hydrodynamic boundary value problem that defines $R_h$.[39,40]

An alternative way to probe the structural aspect of the chain is to calculate the form factor[35] or, the infinite-dilution structure factor, $S(q)$, defined as

$$S(q) = \frac{1}{N}\left\langle \sum_{i=1}^{N-1} \sum_{j=i}^{N} e^{-iq.(r_j - r_i)} \right\rangle, \quad (2)$$

where $q = |\boldsymbol{q}|$ is the wave number, $N$ is the number of atoms, and $\boldsymbol{r_i}$ is the position of the $i^{th}$ atom. This quantity describes correlation of a collection of atoms of a particular type. When plotted as a function of the wavenumber $q$, one expects to find a flat Guinier regime at very small $q$ ($qR_g < 1$), followed by a power-law decay.[47] The value of this power-law ('Porod' exponent) is determined by the chain structure. MDCraft[48] is used to calculate the form factor.

We also looked at the changes in dipolar orientation by calculating the smallest angle, $\theta$, created between the dipole vector $\vec{\mu}$ on the side chain and tangent drawn on the nearest point on the backbone $\vec{b}$. We then plot the probability $P(\phi)$. To examine the dynamics of water trapped inside the polymer mesh and in its vicinity, we have used the Debye-Waller (DW) parameter, $\langle u^2 \rangle$; average mean squared displacement at a caging timescale (1 ps). This quantity is related to the dynamic free volume $\langle u^2 \rangle^{3/2}$ and predictive of mobility at much longer time scales.[49–51] It has been used extensively in the context of glass-forming liquids,[51–53] and here this quantity is used to quantify the local mobility of the solvent molecules about the polymer chains following the procedure described in previous work by Clark *et al*.[32] A smaller value of $\langle u^2 \rangle$ than that of bulk water in the vicinity of the chain indicates a slowing down of the water dynamics in the dynamic hydration layer around the polymer.

To compare the relaxation of the chains and its correlation with the relaxation of the water trapped inside the hydration layer, we have computed the self-intermediate scattering function (SISF),

$$F_S(t) = \frac{1}{N}\sum_{i=1}^{N} \exp(\, i\boldsymbol{q}.\boldsymbol{\delta r_i}), \quad (3)$$

where $\boldsymbol{\delta r_i} = \boldsymbol{r_i}(t) - \boldsymbol{r_i}(0)$, is the displacement of the $i^{th}$ particle after time $t$, and $q = |\boldsymbol{q}|$, is the wave number. On longer time scales, after a short ballistic regime, the SISF can be approximated by a stretched exponential function[22]

$$F_s(t) = A \exp\left[-\left(\frac{t}{\tau}\right)^{\beta}\right], \tag{4}$$

where $A$ is a constant determined by the plateau associated to the cage effect, $\tau$ is the characteristic relaxation time, and $\beta$ is the stretching parameter describing the overall 'shape' of the relaxation function. Particular values of $\beta$ for different models of polymer solutions will be discussed in the following section.

**Results**

Fig. 1(a) and 1(b) show equilibrium snapshots of our model sulfobetaine chain at 320 K in pure aqueous solution and in NaCl solution, respectively, along with its static hydration layer, defined as the water molecules in direct contact with the hydrophilic sites on the chain.

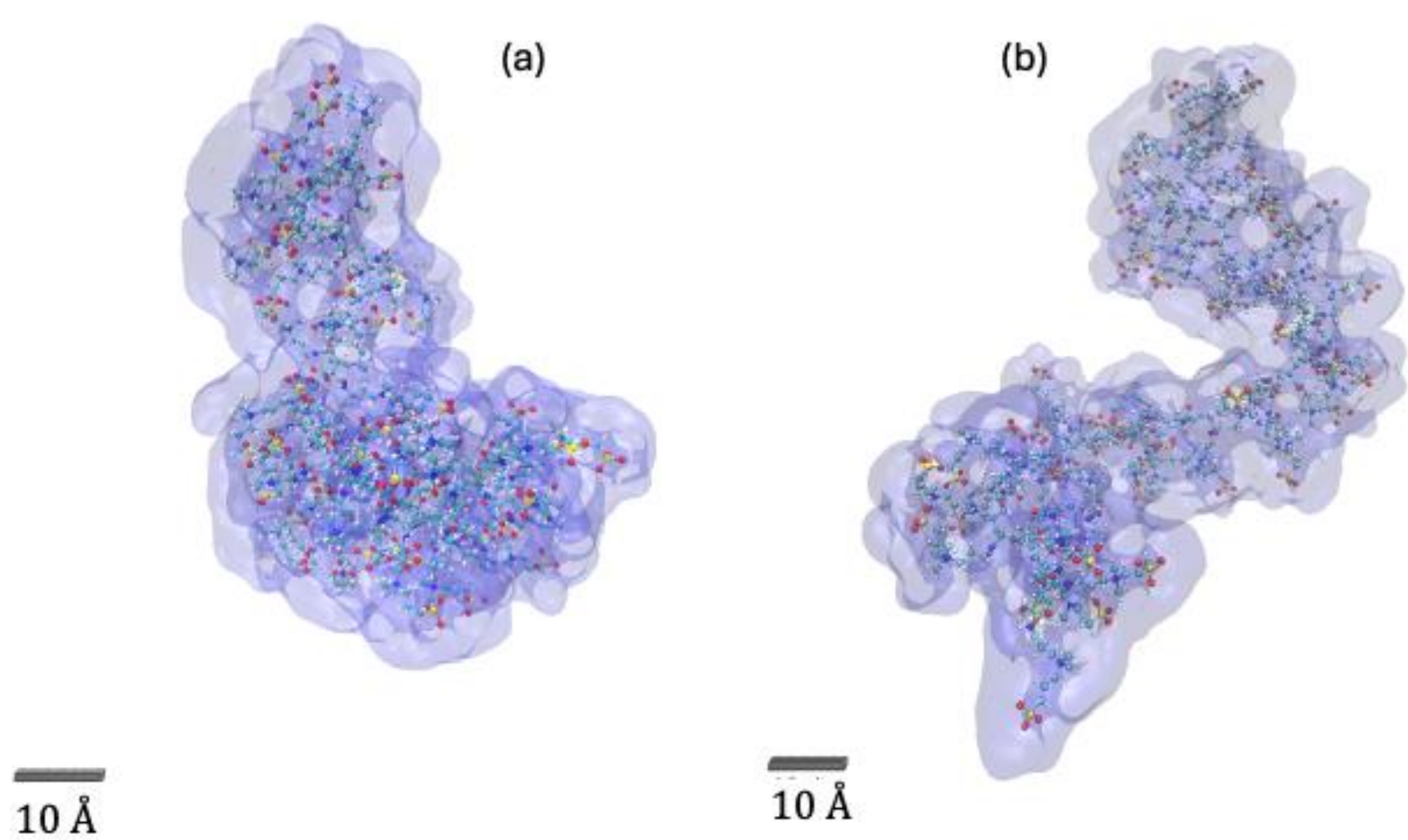

***Figure 1:*** *Visualization pAPAPS chain in equilibrium along with its static hydration layer in (a) salt-free and (b)* $0.5\ mol/L\ NaCl$ *solution, at* $320\ K$.

Fig. 2 shows a comparison of $R_g$ of the chain at different temperatures in salt solution and in salt-free solution. Error bars indicate the standard deviation of the $R_g$ values from the average over the course of production runs. The size of the chain does not show much variation with change in temperature, whereas added salt has a strong effect. For any given temperature, we observe an increase in chain size, a phenomenon that has been termed the *anti-polyelectrolyte* effect.[16,54]

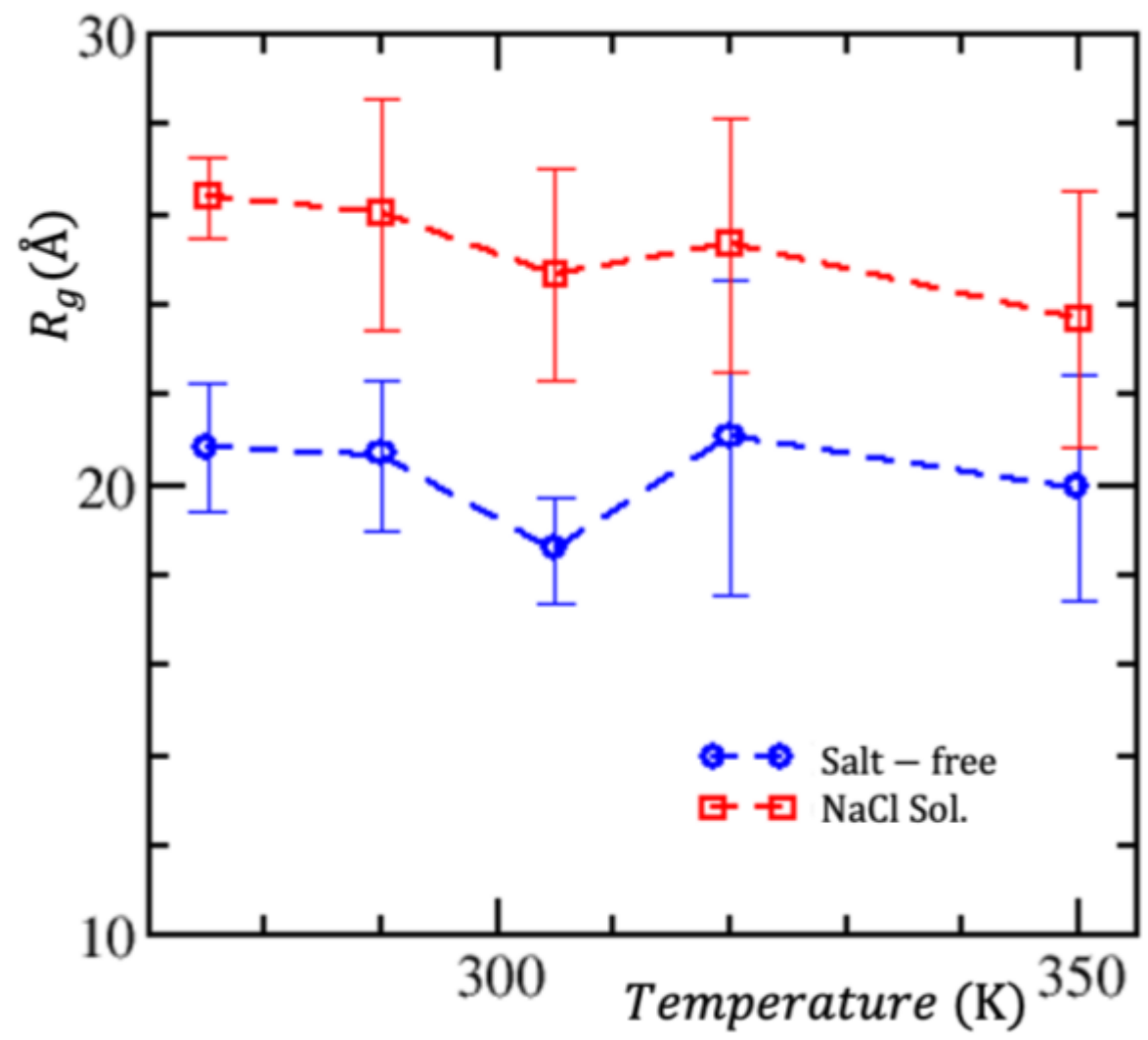

***Figure 2**: Radius of gyration, $R_g$ of the pAPAPS chain as a function of temperature in pure water and $0.5M\ NaCl$ solution is plotted.*

As noted previously, $\psi_h$ is sensitive to the average compactness of polymer chains.[17,55,56] A value of around 1.1 observed (Fig. 3) in the case of the salt-free solution is characteristic of a relatively compact polymer configuration, similar to that found in swollen randomly branched polymers ('lattice animals') and percolation clusters, randomly branched polymers with screened binary excluded volume interactions at the theta point of such polymers[10] and in many-arm star (12-arm) polymers.[37,57] In polymers with a network topology (model single chain nanoparticles) with varying cross-link density, $\psi_h$ has been observed by simulations to vary smoothly with cross-link density over a range of values between that of a sphere to a random coil polymer.[35] With the addition of salt, we observe (Fig. 3) a reduction in magnitude of $\psi_h$, suggesting a less compact polymer configuration, and a weak temperature dependence, absent in salt-free solution. We also observe a counter-intuitive increase of $\psi_h$ from a value comparable to random coil chain to a randomly branched polymer with an increase in temperature in solution with a significant amount of salt (0.5 mol/L NaCl), apparently indicating a progressive 'compactification' of the chain upon heating, a trend superficially similar to that observed in the common water-soluble polymer Poly(N-isopropylacrylamide) or, PNIPAm[58] which exhibits a lower critical temperature phase separation upon heating. Notably, a compactification of PZ chains in salt solutions upon heating, as determined by SANS $R_g$ measurements, has been observed experimentally by Angelopoulos et al.[22] where there is *no evidence of incipient phase separation* upon heating. This striking trend was interpreted by Angelopoulos et al. to arise from an increasing flexibility of the PZ polymers in salt solution upon heating. We examined the persistence length of our model PZ polymer from the initial decay of the bond correlation functions and could not find any conclusive evidence of such a trend. Indeed, the worm-like chain model did not provide an appropriate description of these chains.

We offer another possibility to rationalize this unexpected trend in the $T$ dependence of $R_g$ seen both experimentally[22] and in our simulations, although the universality of this trend in PZ polymers is not yet clear. It is a well-known property of water that its static permittivity decreases appreciably upon heating in a temperature range near room temperature,[59,60] which should have the effect of increasing the strength of dipole interactions within the PZ chains,[61] an effect that should somewhat negate the effect of adding salt, which increase the strength of these self-associations. This general tendency of the dielectric constant to decrease upon heating provides a natural explanation for the tendency for the PZ chains to form more organized self-associated chains upon heating. We note that a trend of this kind has been observed in temperature dependence of the association of charged polymers in polar solvents,[62] where the enhanced association is likewise attributed to the general tendency of the permittivity of the polar solvent to decrease upon heating. We thus tentatively attribute the decrease of $R_g$ with increasing $T$ in Fig. 2 and the appreciable increase of $\psi_h$ upon heating indicated in Fig. 3, when there is appreciable salt in the solution, to this cause rather than a significant change in polymer rigidity with temperature. The tendency of PZ polymers to exhibit upper critical solution phase behavior, i.e., phase separation upon cooling, has been attributed to be in large part to a decrease of the dielectric constant upon heating.[63]

Apart from this unexpected variation of $\psi_h$with temperature in salt solutiions, $\psi_h$ is still well above the value characteristic of random coil polymers, $\psi_h \approx 0.8$[64,65]. Schmidt and Burchard[64] have discussed the fact that experimental estimates of this ratio normally deviate substantially from the often-cited theoretical predictions of the Kirkwood-Riseman theory, which involves a highly approximate configurational pre-averaging approximation[66,67] that is avoided in the ZENO computations (Note that the reciprocal of $\psi_h$ is often encountered in the polymer science literature to discuss the ratio of $R_h/R_g$ and is designated as $\rho$, following a notation introduced by Burchard and coworkers.[68]) Delgado and Schlenoff[10] have further noted the general tendency of experimentally observed values of $\psi_h$ in PZ solutions tend to be larger than found in ordinary synthetic uncharged polymers, as in our simulations, and an attempt to directly visualize model PZ polymers by Jingfeng *et al.*[69] led to the conclusion that the individual chains adopt a roughly spherical configuration in solution. Angelopoulos *et al.*[22] have also attributed the observed compact state of PZ polymers in small-angle neutron scattering measurements to the formation of a self-associated molecular state, in qualitive accord with our simulation observations.

By extension, we note that values of $\psi_h$ significantly larger than random coil estimates have also been extensively observed in intrinsically disordered proteins[70–72] where this polymer compaction effect is less prevalent than in fully unfolded or 'denatured' globular proteins, which apparently have a greater resemblance to random coil polymers.[73,74] Relatively large values of $\psi_h$ have also been reported in high precision measurement of polyelectrolyte complexes of DNA with polyethylenimine,[75] a system of great practical interest in the context of drug delivery applications and estimates of $\psi_h$ near 1 are typically observed in evolutionally-conserved ssRNA molecules of biological interest.[76–78] It is generally appreciated that the diverse biological functions of RNA molecules crucially depend on forming functional structures or single-chain nanoparticles in solution that involve polymer self-association,[79] and it has been broadly appreciated recently that the configurations of these biomolecules can be reasonably described as randomly branched polymers.[80,81] Evidently, the formation of randomly branched polymer structures through intramolecular association and intermolecular complexation is a ubiquitous phenomenon in water-soluble polymers, and the phenomenon seems to be particularly prevalent in intrinsically disordered proteins, RNA and synthetic PZ polymers.

Values of $\psi_h$ close to the hard sphere value of 1.27 are naturally expected in compact globular proteins (see discussion below) so the tendency for $\psi_h$ to increase to a value much larger than random coil polymers when PZ are strongly self-associated in the complete absence of salt. Under these conditions, the PZ conformation appears to be more comparable to globular proteins than intrinsically disordered proteins having a more open and fluctuating structure.

Measurements of $\psi_h$ for a synthetic polymer undergoing a coil-globule transition are instructive in comparison to our PZ simulation results as these measurements illustrate the broad range of values exhibited by $\psi_h$ as solvent conditions are varied and the chain conformation varies from highly swollen to a compact globular state. Precise measurements on the synthetic polymer PNIPAm have established that $\psi_h$ varies from a value near 0.7 in a good solvent to a value around 0.8 near the theta point where $\psi_h$ varies rapidly as the chains begin to collapse to a value near the hard sphere value near 1.27 when the chains are in a fully collapsed state.[58,82] Values of $\psi_h$ even a little larger than 1.27 are observed for a limited temperature range below the theta point where the polymer chains can be described as having a relative dense core and a highly diffuse periphery ('fractal globule'), a behavior observed in high generation dendrimer polymers, polymer grafted nanoparticles[83,84] and non-equilibrium Eden clusters.[36] We note that the term 'fractal globule' has been given to describe chromatin,[85–87] a physically cross-linked form of duplex DNA and no data seem to be available to confirm whether our use of this term conforms to the usage in the case of chromatin. At any rate, a survey of many globular proteins has indicated that $\psi_h$ has an average value of 1.45,[88,89] suggesting that the fractal globule structure, as we define it, rather than a fully collapsed globule is a prevalent structure in globular proteins in their native folded state.

We infer from these observations and our simulations, that the tendency to form self-associated branched polymeric structures, and corresponding $\psi_h$values around 1, can be expected to be a *generic phenomenon* in intrinsically disordered proteins, polyelectrolyte complexes, and PZ polymers, although the extent of the phenomenon must depend appreciably on the charged species in the molecules, their distribution, added salt and temperature, etc. Measurements of $\psi_h$ are expected to be quite useful in the characterization of these important classes of polymers and polymer complexes under different solution conditions as variations of the functional properties of all these polymers can be expected to depend strongly on their conformational structure in solution. We also expect $\psi_h$ to provide a valuable 'reality check' in simulation studies of these classes of polymers.

The simple random coil paradigm of flexible polymers is evidently not sufficient in these important classes of polymers and further research is required on this type of intramolecular self-assembly process and its impact on the solution properties of these polymers. We next move to considering the scattering properties of our model PZ polymers.

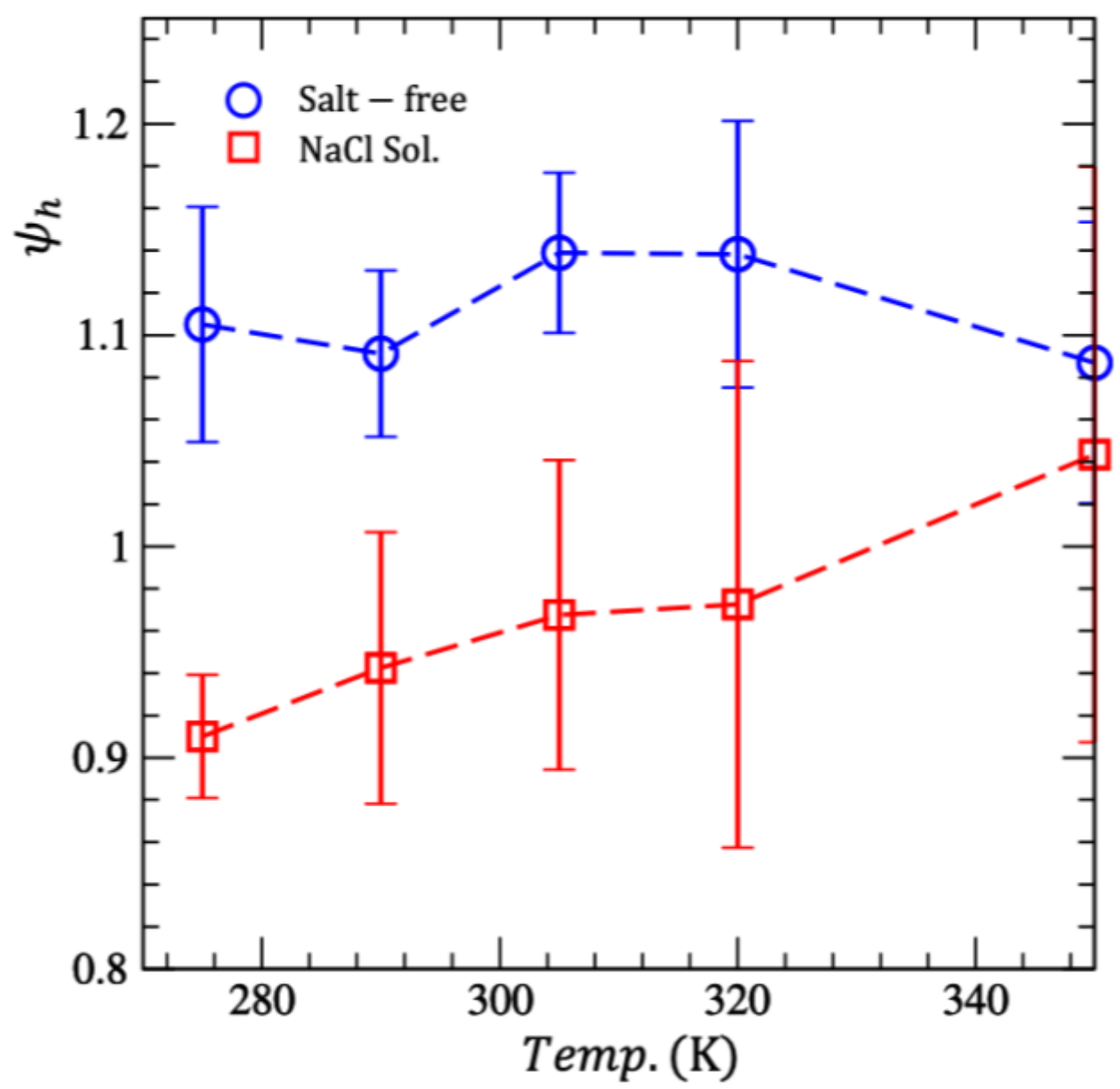

***Figure 3:*** *The quantity $\psi_h = R_h/R_g$ for the model polyzwitterion chain is shown as function of temperature in pure water and $0.5\ mol/L\ NaCl$ solution. Standard deviation in $\psi_h$ from the average value calculated from the production runs are presented as error bars.*

The intramolecular form factor calculated for the backbone carbon atoms where the sidechains are connected is shown in Fig. 4. The power-law values are calculated from the $S(q)$ data following the Guinier plateau and plotted in the inset. Uncertainties in the fit are smaller than the symbol sizes and hence not shown for clarity. Power-law scaling is observed with the Porod indices $3, 2, 1.67$ and $1$. A value of $3$ is indicative of a compact configuration and a value of $1$ indicates rod-like configurations[21]. A value near 2 is characteristic of both high mass random coil and swollen randomly branched polymers,[35] while a value near $5/3 \approx 1.67$ is expected for swollen linear chains[90] and an exponent near 5/2 should be expected for high mass randomly branched polymers under theta conditions.[25] It is generally difficult to ascribe the conformational state of polymers based on estimation of this apparent exponent, but this information can be useful in conjunction with other observations. A significant change in the Porod exponent can be appreciated from these plots due to added salt, irrespective of the temperature, indicating an appreciable change in polymer back-bone conformation. We next turn to investigating whether this conformational change has anything to do with a change in dipolar orientation in the side-groups that has been suggested theoretically.[5]

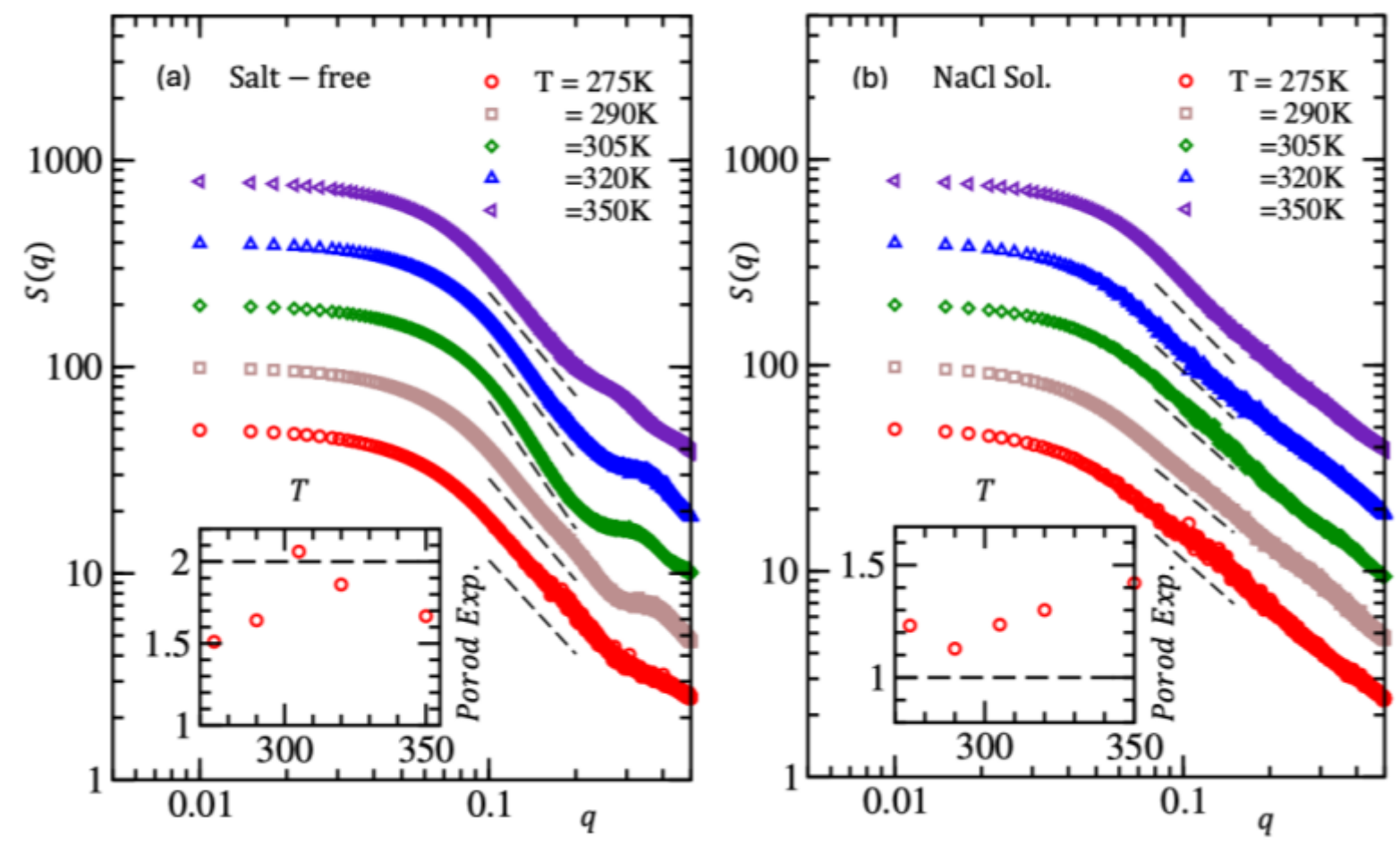

***Figure 4:*** *Form factor of the backbone carbon atoms, $S(q)$ are plotted as a function of wavenumber $q$ for the chain in (a) pure water and (b) $0.5\ mol/L$ NaCl solution. The curves are vertically shifted along the y-axis for clarity by a successive factor of 2. Inset shows the estimated Power-law values.*

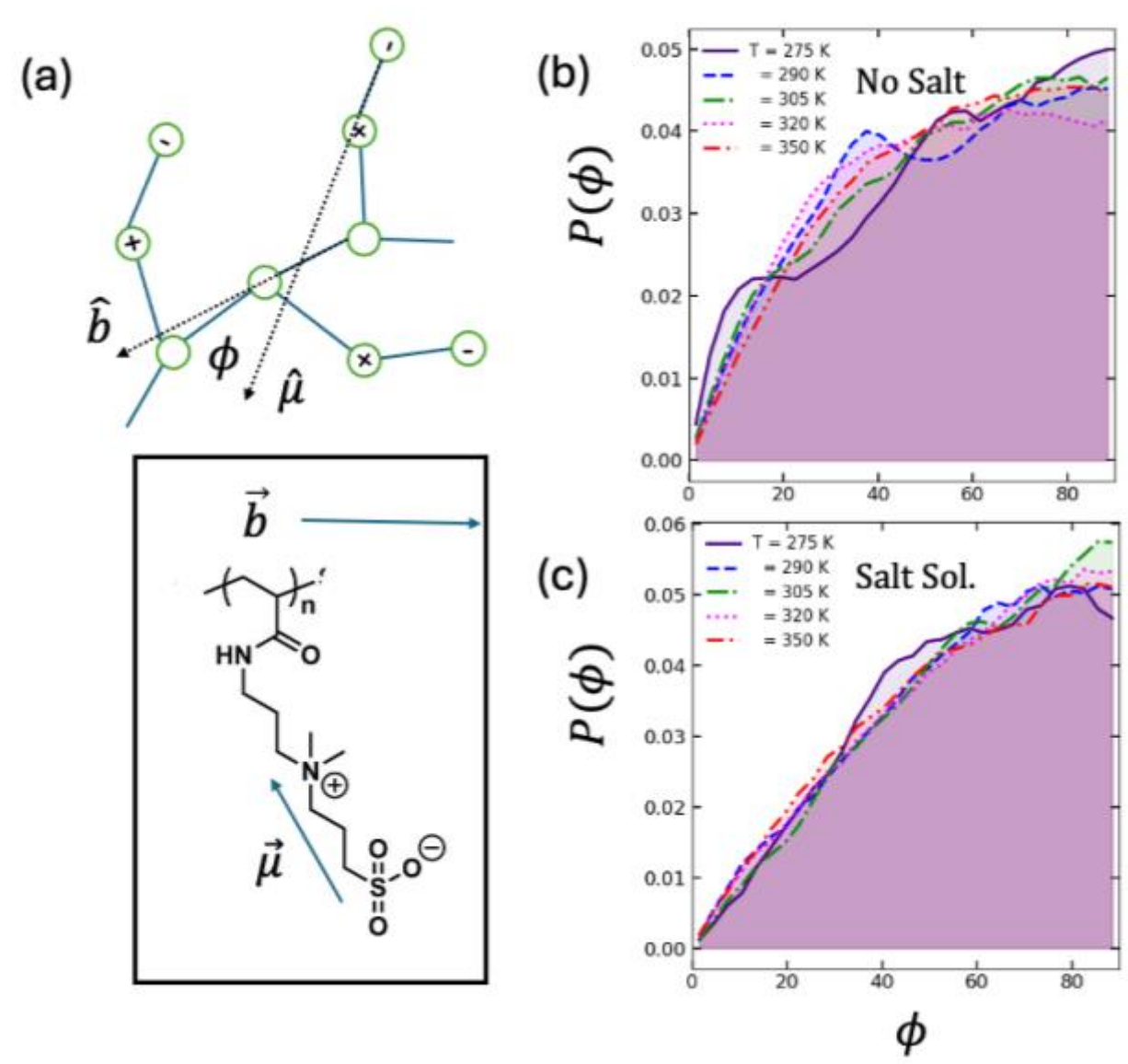

***Figure 5****: (a) Schematic diagram to illustrate how the angles, $\phi$, are calculated. (b) Probability, $P(\phi)$, as a function of $\theta$ is plotted for the chain in pure water. The x-axis is between $0$ and $90°$ as the smallest angle created between the dipole and backbone vector is recorded. (c) same as (b) but for NaCl solution.*

Both temperature and salt are expected to have some effect on the side-chain orientations. We looked at the orientations of the dipoles directed from negatively charge center to positive charged center on the sidechains. Fig 5(a) shows schematically how the smallest angle, $\phi$, between backbone vector $\vec{b}$ and dipole vector $\vec{\mu}$ are calculated, along with the chemical structure of a monomer unit. To calculate $\vec{b}$, we have considered only those carbon atoms on the backbone where sidechains are connected. Fig. 5 (b) and 5 (c) show how the probability, $P(\phi)$ changes with change in temperature, in salt-free and NaCl solution, respectively. In salt-free solution, we see parallel/anti-parallel arrangement of dipoles for temperature $305$ K and above, while for temperatures lower than $305$ K, dipoles tend to orient along the backbones, giving rise to the peaks at $\phi < 90°$. This is surprising considering the fact that temperature does not affect the overall size of the chain, evident from the radius of gyration data. When salt is added, we observe the presence of such a peak, although it is rather weak and only at the lowest temperature considered, i.e., $275$ K, to emphasize the effect.

To better understand the geometry of the chain, we quantified the distribution of the ionic functional groups as distance perpendicular to the backbone. For both nitrogen atoms in the quaternary ammonium group, we shall simply refer to them as 'ammonium' hereafter, and oxygen atoms in the sulfonate group, we see a non-Gaussian distribution around the chain. The distribution of the nitrogen atoms is much narrower than the oxygen on the sulfonate group. Fig. 6 shows such distribution at two different temperatures, $275$ K and $320$ K, for pure aqueous solution, and at $320$ K for NaCl solution. With variation in temperature, and salt concentration, we observe a change in height and position of the peaks, but the width of the distribution remains the same. These plots provide a rough idea about the effective diameter of the PZ chain. The range over which the pendent groups are distributed is critical to understand the extent of the hydration layer. The nitrogen in the ammonium group can be found roughly between 3 Å and 8 Å while the oxygen of the sulfonate group ranges from $3$ Å to $12$ Å. The furthest distance up to which the functional groups are distributed from the backbone of the chain, can be identified as the effective cross-sectional radius of the chain, $\approx 12$ Å for our PZ chain.

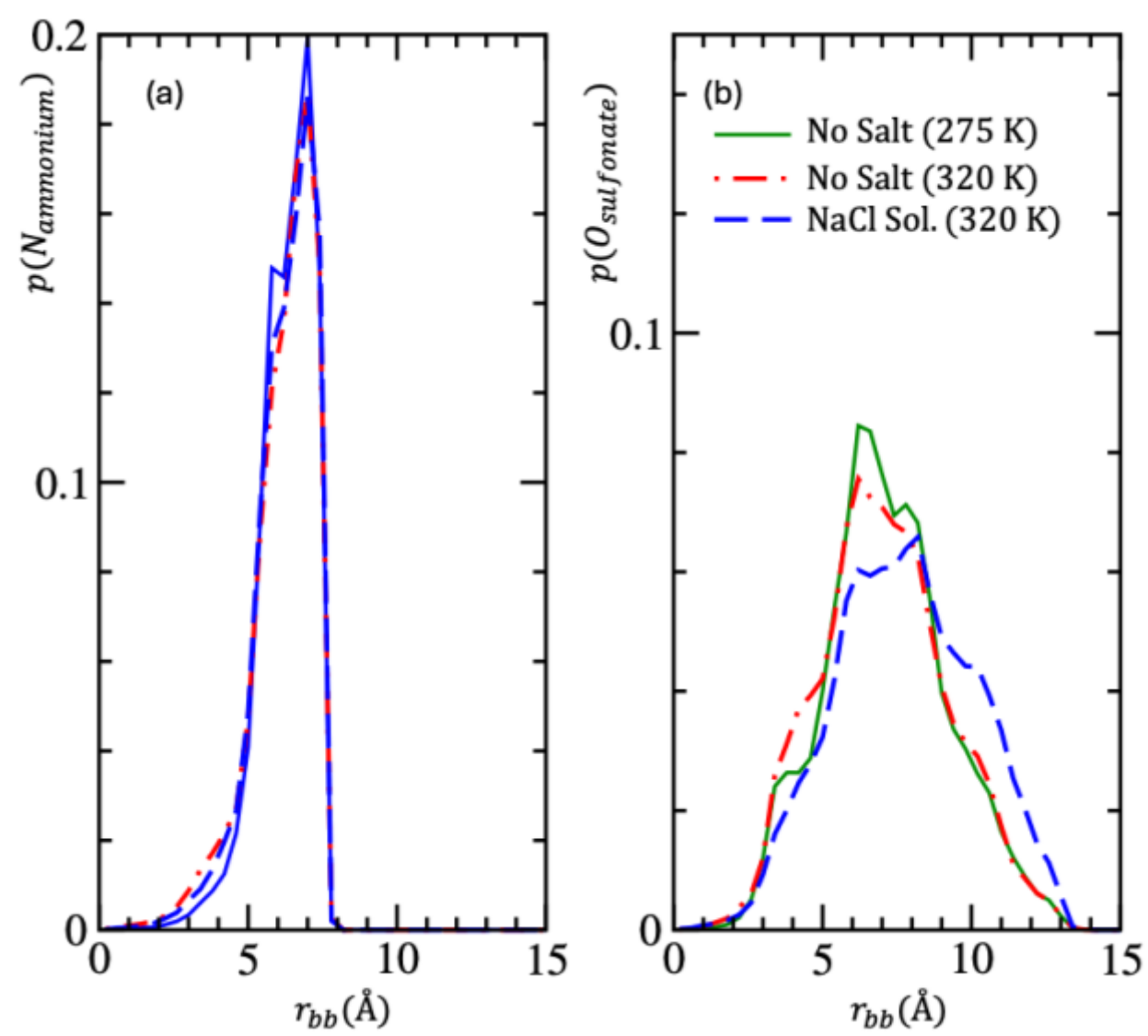

***Figure 6**: Distribution of the ionic functional groups as a function of distance, normal to the backbone vector. (a) shows distribution of Nitrogen present in the quaternary ammonium group. (b) distribution of oxygen in the Sulfonate group.*

Different experimental techniques have been introduced over the past few decades to study the extent of dynamic hydration layers wrapping the macromolecules[91,92]. In simulations, a geometric cutoff can be used to separate hydration water from the bulk. Here, following an earlier study[32], we have used the local mobility metric, $\langle u^2 \rangle$, the average mean squared displacement at 1ps, as a tool to probe the water dynamics. In Fig. 7, we show plots of $\langle u^2 \rangle$, relative to its value at bulk, $\langle u_0^2 \rangle$, as a function of distance from the backbone. The quantity increases as the distance increase and goes to 1, corresponding to 'bulk' limit, at a distance $\approx 20$ Å from the backbone and approximately 60 % drop in solvent mobility close to backbone, compared to bulk. The yellow shaded region roughly indicates where the functional groups can be found. The surface of the chain at $\approx 12$ Å marks the boundary for charged pendent groups and can be imagined as the furthest distance from the backbone at which the atoms on the chain can be in direct contact with the solvent. In salt-free solution, we see a strong temperature dependent water dynamics in that region. As temperature is increased the mobility of hydration water, relative to bulk, reduces, indicating a stronger hydration shell. This can be connected to the dipolar orientation of the side-chains. In 0.5 mol/L NaCl solution, this temperature dependence is not so prominent, and we can observe a nominal increase in mobility at 275 K. Inset of Fig. 7(b) shows the variation of $\langle u_0^2 \rangle$ of water with temperature and we can see a reduction in mobility due to added salt.

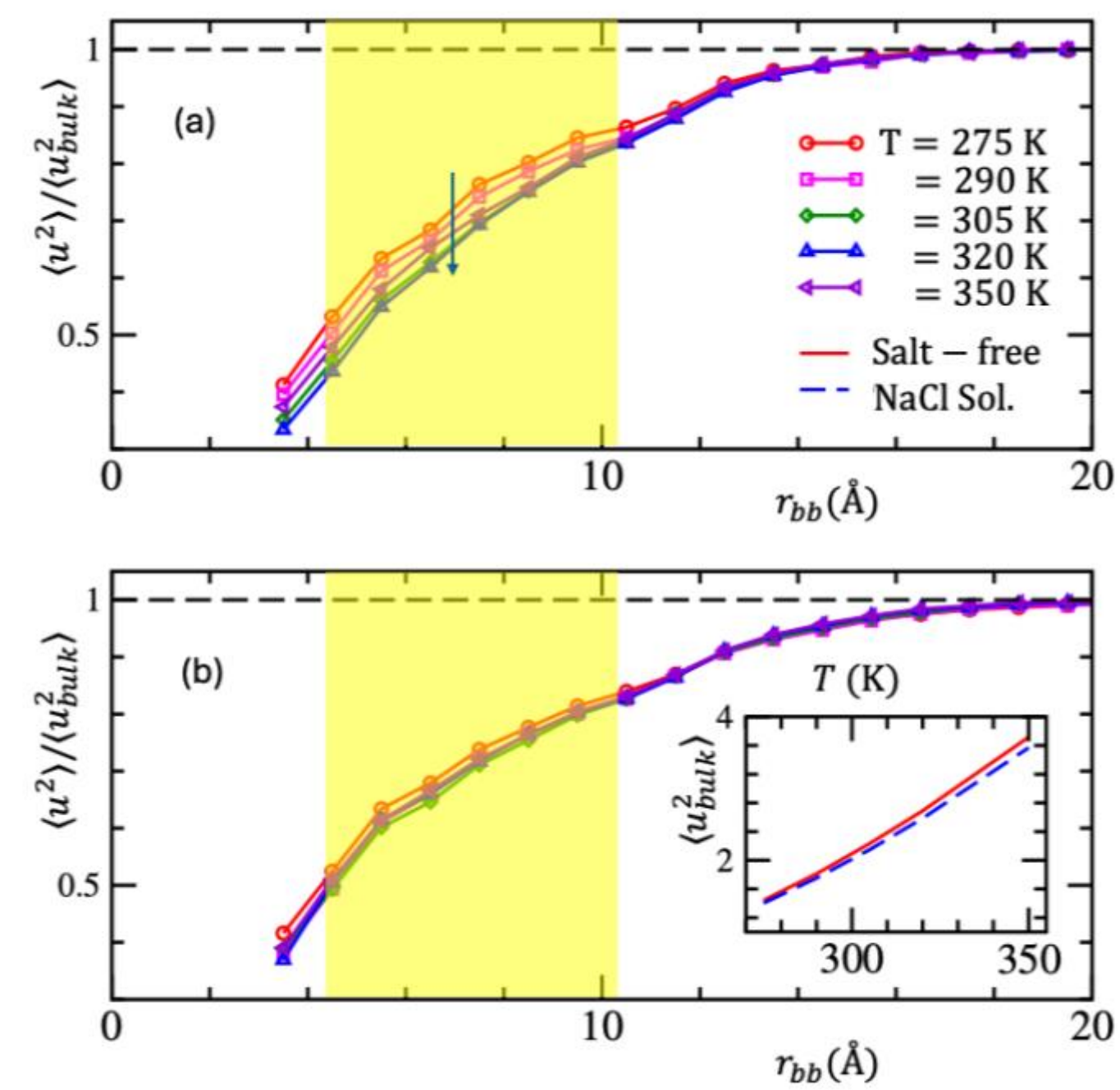

***Figure 7****: DW parameter, $\langle u^2\rangle$, of water is plotted as a function of distance from the backbone, relative to its value at bulk in (a) salt-free solution and (b) $0.5\ mol/L\ NaCl$ solution. Inset shows the values of $\langle u^2\rangle$ of water in the bulk solution. The arrow indicates the shift in relative Debye-Waller parameter with increasing temperature.*

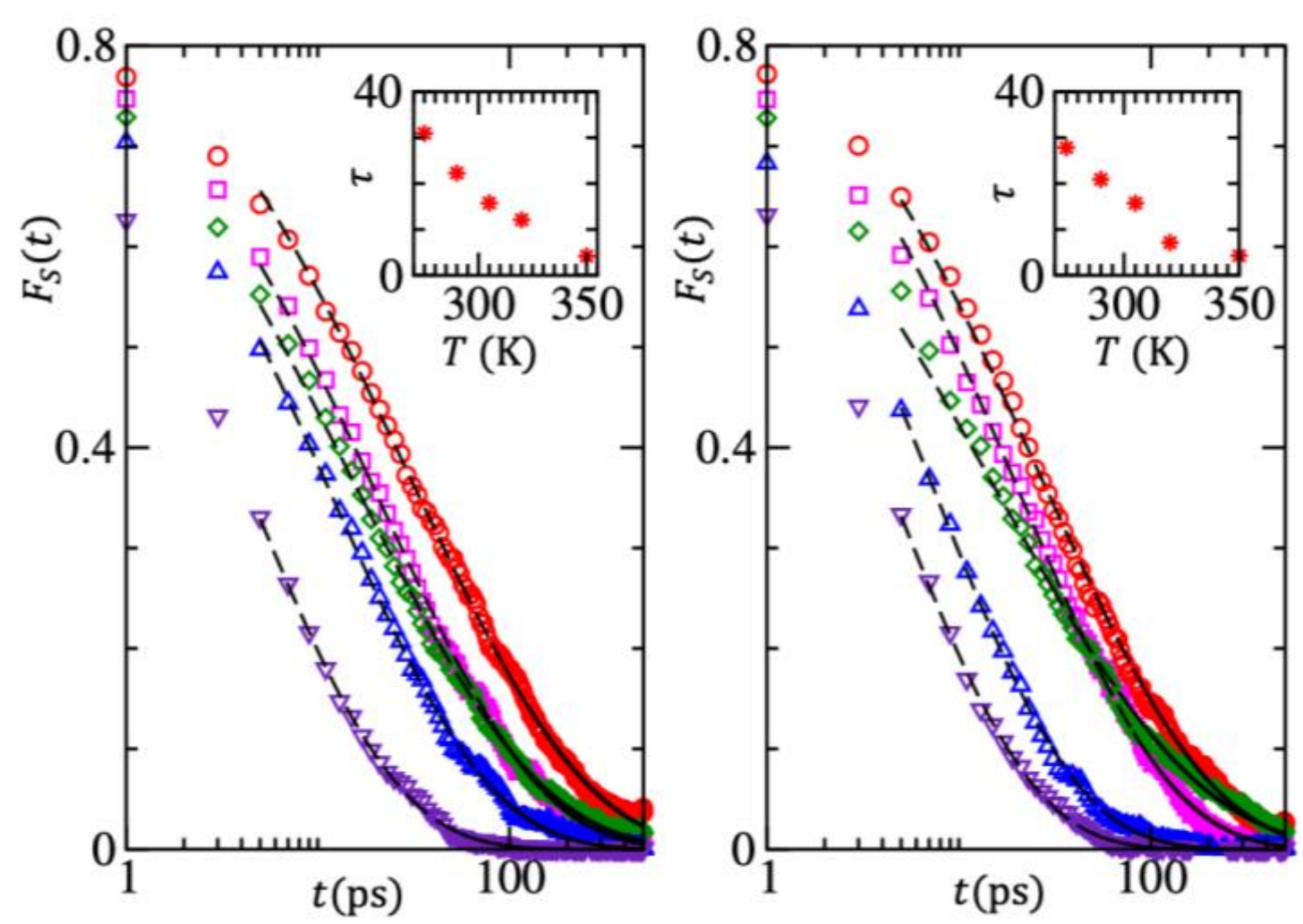

***Figure 8****: SISF, $F_S(t)$, is plotted as a function of time for the atoms on PZ chain in salt-free (left) and 0.5 mol/L NaCl solution (right) with symbols. The dashed lines are fit to Eq. (4). Inset shows the change in characteristic timescale with change in temperature.*

Fig. 8 shows the SISF data calculated for the PZ chain is salt-free, as well as NaCl solution. We have fit our data with the stretched exponential function described in Eq. (4). In both the cases, we see an increase in characteristic time with decrease in temperature. The stretching parameter is found to have an empirical value between $0.4$ and $0.55$ (with around 10% uncertainty as estimated by systematically varying $\beta$ over a range and checking the quality of the fit). As a point of comparison, $\beta$ is predicted theoretically[93] to equal $1/2$ and $2/3$ for ideal random coil polymers, without and subject to very strong hydrodynamic interactions, respectively. The non-draining limit of strong hydrodynamic interactions is expected to be a good approximation in dilute solutions of long flexible polymers[94]**,** while the Rouse model provides an approximate model for unentangled polymers in the melt where hydrodynamic interactions are apparently 'screened' because of the strong interpolymer excluded volume interactions.[95,96] The exponent has a clear interpretation in the coarse-grained Rouse and Zimm models of polymer dynamics in terms of the mean square displacement $\langle r^2 \rangle$ of the segments within the polymer with time $\langle r^2 \rangle \sim t^{\beta}$ over scales less than the overall polymer size, the motions probed by the intermediate scattering at high wavevector $\boldsymbol{q}$ where the interior of the polymer is probed in the scattering measurement.[97] This physical interpretation of $\beta$ provides a framework for interpreting changes in this quantity with changes in thermodynamic conditions and the topological structure of the polymer.

Martin and coworkers[98] have emphasized that $\beta = 2/3$ should be a rather 'universal' value of $\beta$ for linear polymers in either good (toluene) or theta (cyclohexane) solvents should be rather insensitive to solvent quality *when the polymer is at least somewhat soluble.* A number of studies have argued that $\beta$ should have a somewhat modified value $\beta = 3/4$ in solutions of semi-flexible polymers where hydrodynamic interactions within the polymer are weaker and the polymer chains thus more hydrodynamically 'permeable',[33] and, by extension, it seems plausible that $\beta$ might become somewhat smaller than $2/3$ for PZ and other polymers exhibiting significant self-association since that the polymers might then acquires a gel-like dynamics in which the effects of hydrodynamic interactions are suppressed as in the case of concentrated polymer solutions and melts. If this hypothesis is correct, then globular proteins and other molecules exhibiting self or mutual complexation might be expected to follow a similar trend in the intermediate scattering function as we have seen for our PZ polymer.

In line with this reasoning, neutron spin-echo measurements on model proteins (hemoglobin and myoglobin) in dilute solution indicated an apparent value[99] of $\beta$ near $0.4$, consistent with our estimate of $\beta$ for our model PZ polymer. Reuveni *et al.*[97] have introduced a quantitative theory of the self-intermediate scattering function for polymers having this type of elasticity aimed at describing relaxation in dilute protein solutions where values of $\beta$ consistent with values noted above are predicted. Interestingly, Milchev and Binder[100] also observed anomalous diffusion in collapsed flexible polymers below their theta point with a significantly lower $\beta$ values than polymers in good and marginal solvents above the theta point where $\beta$ was found to be strongly temperature dependent so that the anomalous diffusion exponent is apparently tunable in the collapsed state. We also note that $\beta$ values in the range 0.4 to 0.42 have been reported[101–103] in simulations of fractal globules in solution, an observation consistent with molecular tracking measurements of biological complexes of interest such as telomeres complexes where was found to near $\beta = 0.4$. Substantial evidence would then seem to indicate that $\beta$ is not really universal for polymer solutions. The emergence of rigidity of the polymers through intermolecular association[104] or cross-linking would appear to suppress hydrodynamic effects on the internal dynamics of the polymer, making the polymer dynamics sensitive to the fractal structure.[97,105] This subtle phenomenon of wide potential practical significance evidently deserves further study.

As a practical matter, the empirical exponent $\beta$ provides an economical way to quantify the change in the shape in the intermediate scattering function and other relaxation functions in our PZ system which is useful for making comparisons with the dynamics of flexible polymer solutions in marginal and good solvents. The assumed stretched exponential functional form for $F_S(t)$ also allows us to determine the structural relaxation time, which we have extracted from the fits of $F_S(t)$ for our PZ polymer to the stretched exponential as a function of time. The fitted values of the relaxation time are shown in the inset to Fig. 8, where an appreciable slowing down of the chain dynamics upon colling is observed. We next show that a parallel slowing down occurs in the hydration water around the PZ chain and in the PZ side-groups, although the slowing in the dynamics within the dynamic hydration layer shows some distinctive features.

Existing literature suggests that the hydration water plays a critical role in structural stability and dynamics of polymer chains in aqueous solution[91]. Many previous studies indicate that the proteins and the water molecules inside their hydration shell mutually influence each other, resulting in a strongly coupled dynamics[106,107] and it has even been claimed that the protein dynamics is "slaved" to the hydration water dynamics. Existing literature also suggests that this 'coupling' of the side-group dynamics to the water dynamics is residue-specific[108], as well as temperature sensitive[109]. To briefly probe whether this kind of coupling between polymer and water dynamics is occurring in our PZ system, we calculated the relaxation times of PZ functional groups in the side-chains. In particular, we calculate the relaxation time from the intermediate scattering function for the oxygen atoms in the sulfonate groups and nitrogen atoms within the ammonium groups and compare these times with the relaxation times for the water molecules inside the static hydration layer. Fig. 9 (a, b) and 9 (c) show changes in these relaxation times, $\tau$, relative to their value at $350$ K, in salt-free solution, $\tau_0$, with change in temperature for pendent groups and water molecules residing in the static side-chain hydration layer, i.e., a solvent molecules in direct contact with the pendent groups. We see that all these relative relaxation times show a similar overall trend: a significant increase in $\tau$ upon lowering temperature, indicating that both the side-group and hydration layer relaxation times grow in a somewhat parallel fashion so that the side-group dynamics is apparently coupled to the hydration water. However, the dynamics of the side-groups are not completely 'slaved' to the hydration dynamics, as these relaxation times show some distinctive variation with temperature. There is a weak non-monotonic temperature dependence of the pendent group relaxation time, suggesting that a conformational change of the chain associated with the dipolar reorientation in the side-groups also contributes to the dynamics of the side-groups. Added salt evidently also alters the intensity in the change of the side-chain relaxation with temperature. Moreover, the dynamics of water in the hydration shell become slightly faster with the addition of salt, as evidenced by the reduction in relaxation times shown in Fig. 9 (c); this is completely different from the bulk water behavior, shown in the inset of Fig. 7 (b). Experimental studies have indicated that the polyzwitterion and charged polymer layers can lead to large reductions of surface friction where this phenomenon has been attributed to the effect of the hydration layer[110] and the friction reduction in these polyzwitterion zwitterion layers, while being roughly insensitive to salt concentration, has been found to be influenced somewhat by specific ionic effects that follow the Hoffmeister series[111,112]. Given the practical importance of this phenomenon in connection with lubrication processes in living systems, further research should be made on this coupling of the hydration layer and polymer side-chain dynamics.

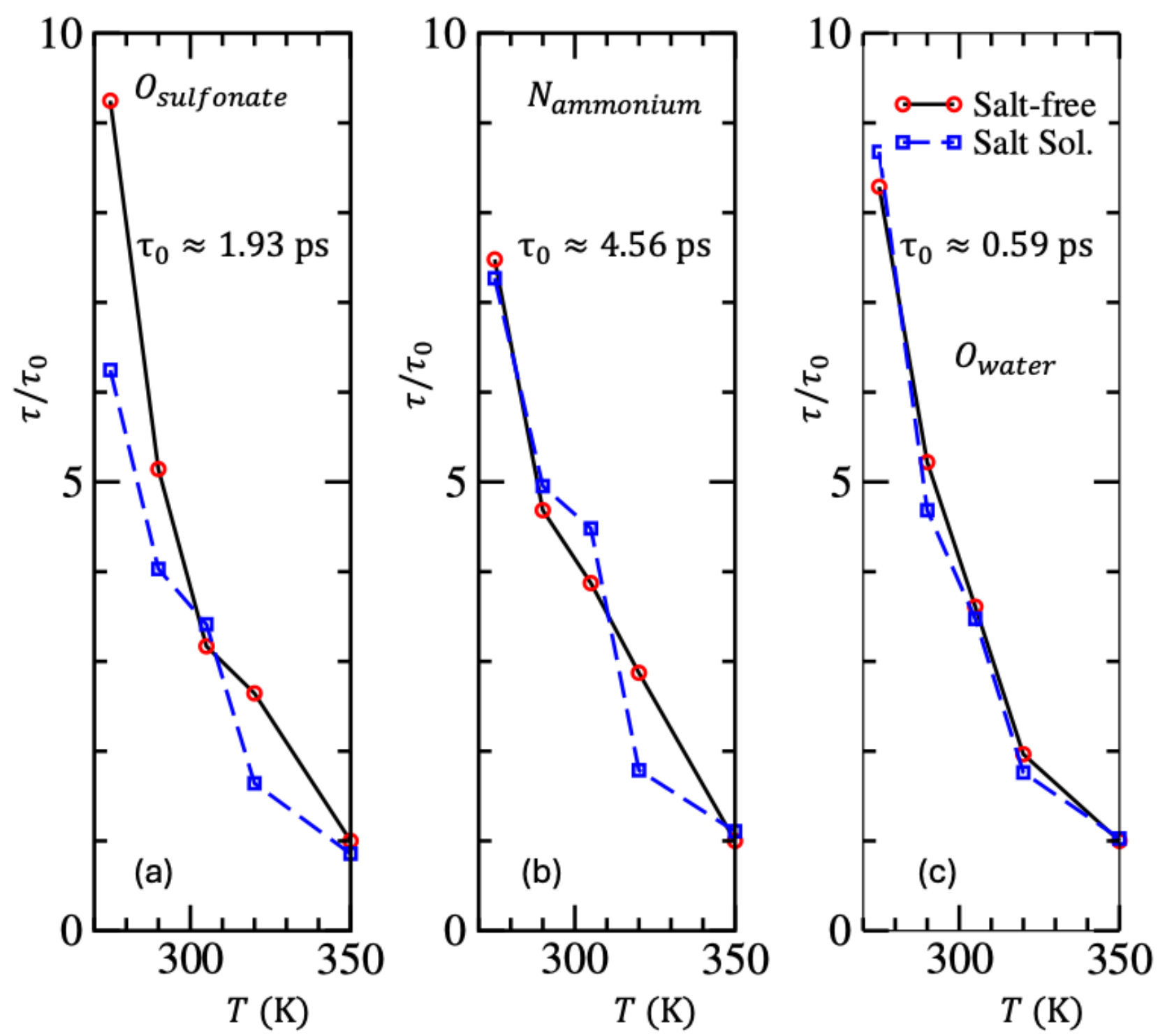

***Figure 9**: Relaxation times, $\tau$, relative to their value in salt-free solution at $350K, \tau_0$, calculated from self-intermediate scattering function, $F_s(t)$, for (a) oxygen in sulfonate group, (b) nitrogen in ammonium is shown. Solid lines with circles and dashed line with squares show the values for salt-free and $0.5mol/L$ NaCl solutions, respectively. (c) and (d) shows the same quantity calculated for water (oxygen atom) in static hydration layer and in bulk, respectively.*

## Conclusion

We performed exploratory molecular dynamics simulations on a model polyzwitterion polymer with a physically realistic water model in which ion-solvent interactions have recently been shown to be consistent with observed ion-specific changes in water mobility that strongly correlate with the Hofmeister series.[113] This molecular model is anticipated to provide a sound foundation for studying ion and polymer hydration in this important class of polymers that greatly influence polymer configurational properties and the dynamics of these polymers.

Although previous molecular dynamics simulation studies of polyzwitterion polymers with explicit solvent are extremely limited, there is ample experimental evidence that the solution properties of these polymers are rather different from both uncharged synthetic flexible polymers in organic solvents and polyelectrolytes. For example, measurements have shown that mean dimensions of flexible polyzwitterion polymers tend to be rather insensitive to temperature so that coil-globule transition as solvent quality is varied does not arise in these polymers and the addition of salt can cause the polymers to expand in dimension when salt is added ('anti-polyelectrolyte effect'). The high biocompatibility of these polymers with biological materials, antibiofouling characteristics, low friction coefficient and other properties of polyzwitterion polymers all point to these polymers being physically similar to biopolymers having a similar amphiphilic character such as intrinsically disordered proteins, RNA and polyelectrolyte complexes of polymers of opposite charge valence involved in the formation of diverse 'coacervate' biological materials. We draw upon this rich phenomenology to ask questions about the nature of polyzwitterion conformation, polymer hydration and the influence of salt on these polymers.

Our simulations confirm the widespread observation that the conformation of polyzwitterion polymers do not show the characteristic strong dependence of the chain size, as determined by chain radius of gyration and hydrodynamic radius, when temperature is varied. We also find that the value of the hydrodynamic penetration function, $\psi_h = R_h/R_g$, tends to have a value much larger than a value representative of a non-associating flexible linear polymer. In particular, we find $\psi_h \approx 1.1$, a value of this ratio typical for randomly branched polymers near their theta point[35,106,114] where the second virial coefficient vanishes. Moreover, $\psi_h$ is found to be remarkably insensitive to changes in temperature and to salt, in comparison to variations with salt seen polyelectrolytes. These trends are instead reminiscent of globular proteins where the polymer conformations are relatively compact and insensitive to thermodynamic conditions, and which exhibit a theta point defined by a vanishing second osmotic virial coefficient,[115] just as observed in measurements on polyzwitterion polymers.[21] The formation of single-chain nanoparticle or 'nanogel' domains in solution resolves this paradoxical finding in polyzwitterion polymers. With the addition of a moderately high amount of salt ($0.5$ mol/L NaCl) to the polyzwitterion polymer solution, we observe the polyzwitterion change to a significantly less compact conformation, but still more compact than ideal random coil polymers in theta or good solvents, as quantified by the observed range of $\psi_h$ values seen in our simulations which are consistent with a conformation similar to randomly branched polymers in theta or good solvents. Remarkably similar trends are observed in intrinsically disordered protein solutions[70] and evolutionally-conserved ssRNA molecules of biological interest,[76–78] and polyelectrolyte complexes of DNA with polyethylenimine,[75] a system of practical interest in the context of immunotherapy and drug delivery applications,[75,116] along with other systems exhibiting intermolecular association. Moreover, we observe the commonly reported 'anti-polyelectrolyte effect' in our model polyzwitterion polymer with the addition of salt, i.e., the chain dimensions increase with the addition of salt rather than decrease as normally reported for polyelectrolytes. The addition of salt evidently somewhat disrupts the self-associated structure of the PZ molecules, formed in salt-free solution, to increase solubility, a trend that is clearly reflected in the magnitude of $\psi_h$. This trend is also characteristic of polyelectrolyte complexes, which fully disassociate into separated polyelectrolyte chains when sufficient salt is added to the solution.

We expect $\psi_h$ measurements to be a valuable tool for characterizing simulated conformational ensembles of intrinsically disordered proteins,[117] RNA, polyelectrolyte complexes and even chromatin DNA complexes with proteins that involve significant intermolecular association and formation of branched polymer structures. This is a topic of great scientific interest given the difficulties of characterizing this class of polymers by methods conventionally used in characterizing globular proteins.

In addition to characterization of the conformational properties of our polyzwitterion polymer, we focus on the nature of structure and dynamics of their hydration layer. The side-chains exhibit a major reorganization of dipolar orientation below 305 K, suggested previously theoretically; this change being reflected by a modest change in the overall dimensions of the polyzwitterion polymer. It seems likely that this change might be responsible for the change of the sign of the second virial coefficient in the temperature range, but this will require further investigation to calculate the second virial coefficient as a function of temperature in our polyzwitterion solution. This subtle change in structure of the polyzwitterion polymer is accompanied by a change in the dynamics of the water molecules within the hydration layer of the polyzwitterion polymer side-groups, which are rather strongly coupled to the dynamics of the polyzwitterion side-groups. In salt solutions, this transition in dipolar orientation occurs at a much lower temperature. For this study, we only considered simulations for five different temperatures covering a wide range, and it will be interesting to repeat this exercise for more temperatures and with different salt concentrations to better understand this transition and its physical consequences for the phase stability of our polyzwitterion solutions. Evidence for a conformational change in the same temperature range as the dipolar reorientation transition is observed and provides some insight into the occurrence of a subtle conformational change, but this type of data conveys limited information about the exact nature of the conformational transition induced by the dipolar reorientation transition. We also visualize the hydration layer extending out from the polyzwitterion polymer back-bone, drawing on ideas drawn from glass-forming liquids to quantify the mobility gradient in this extended layer that 'dresses' these polymers. Curiously, the spatial extent of the polyzwitterion hydration layer is found to be generally on the order of a nm, and this scale is counter-intuitively rather insensitive to the addition of salt, even at a moderately high salt concentration. This trend has been observed previously in other model water soluble polymers and globular proteins[118] not prone to self-association.[113]

**Supplementary Material**

The supplementary material of this article contains detailed descriptions of the process to calculate the hydrodynamic radii using the software ZENO[46] from LAMMPS[29] generated trajectory files.

**Acknowledgement**

Computations were performed using the high-performance computing facility provided by the National Institute of Standards and Technology (NIST). The authors acknowledge funding from NIST award 70NANB24H025. Certain equipment, instruments, software, or materials are identified in this paper in order to specify the experimental procedure adequately. Such identification is not intended to imply recommendation or endorsement of any product or service by NIST, nor is it intended to imply that the materials or equipment identified are necessarily the best available for the purpose.